\documentclass[a4paper,twoside,notoc,11pt]{JHEP3}
\pdfoutput=1
\usepackage{epsfig,multicol,slashed}
\usepackage{delarray,amsmath,bbm}

\newcommand{\nn}{\nonumber}
\newcommand{\beq} {\begin{equation}}
\newcommand{\eeq} {\end{equation}}
\newcommand{\beqa} {\begin{eqnarray}}
\newcommand{\eeqa} {\end{eqnarray}}

\newcommand{\ie}{{\it i.e.}}
\newcommand{\eg}{{\it e.g.}}

\newcommand{\wrt}{{\it wrt.\ }}
\newcommand{\rhs}{{\it rhs.\ }}
\newcommand{\lhs}{{\it lhs.\ }}

\newcommand{\as}{\alpha_s}

\newcommand{\la}{{\Lambda}}
\newcommand{\chat}{\hat{\bs{\ell}}}

\newcommand{\ieps}{i\varepsilon}

\newcommand{\veps}{\varepsilon}
\newcommand{\order}[1]{${\cal O}\left(#1 \right)$}
\newcommand{\morder}[1]{{\cal O}\left(#1 \right)}
\newcommand{\eq}[1]{(\ref{#1})}
\newcommand{\fig}[1]{Fig.~\ref{#1}}
\newcommand{\pit}{(2\pi)^3}
\newcommand{\dpv}{\frac{d^3\pv}{(2\pi)^3}}
\newcommand{\dpve}{\frac{d^3\pv}{(2\pi)^3\,2E_{p}}}
\newcommand{\rket}{\ket{0}_{R}}
\newcommand{\rbra}{{_{R}\bra{0}}}

\newcommand{\inv}[1]{\frac{1}{#1}}
\newcommand{\ket}[1]{\vert{#1}\rangle}
\newcommand{\bra}[1]{\langle{#1}\vert}

\newcommand{\com}[2]{\left[{#1},{#2}\right]}
\newcommand{\acom}[2]{\left\{{#1},{#2}\right\}}

\newcommand{\bs}[1]{\boldsymbol{#1}}
\newcommand{\sla}{\slashed}

\newcommand{\ocal}{\mathcal{O}}
\newcommand{\dcal}{\mathcal{D}}
\newcommand{\lcal}{\mathcal{L}}

\newcommand{\qu}{{\rm q}}
\newcommand{\qb}{{\rm\bar q}}

\newcommand{\xv}{\bs{x}}
\newcommand{\yv}{\bs{y}}
\newcommand{\kv}{\bs{k}}
\newcommand{\pv}{\bs{p}}

\newcommand{\gv}{\bs{\gamma}}
\newcommand{\sv}{\bs{\sigma}}
\newcommand{\nv}{\bs{\nabla}}
\newcommand{\av}{\bs{A}}

\newcommand{\halft}{{\textstyle \frac{1}{2}}}
\newcommand{\quart}{{\textstyle \frac{1}{4}}}
\newcommand{\gsim}{\buildrel > \over {_\sim}}

\title{\center{Bound states at lowest order in $\hbar$}}
\author{Paul Hoyer\\
              Department of Physics and Helsinki Institute of
              Physics\\
              \ POB 64, FIN-00014 University of Helsinki, Finland \\
              }

\abstract{Bound states poles in scattering amplitudes are generated by the divergence of the perturbative series due to enhanced Coulomb scattering near thresholds. This suggests to organize bound state calculations according to an expansion in $\hbar$, \ie, in the number of loops. I study QED and QCD bound states at lowest order in $\hbar$, which are analogous to Born amplitudes. The absence of loops allows the use of retarded boundary conditions where particles only propagate forward in time, making a hamiltonian approach feasible. The instantaneous $A^0$ field is determined by the equations of motion separately for each Fock component of the bound state. The field equations are compatible with a linear $A^0$ potential as a homogeneous, non-perturbative solution. Stationarity of the action determines the direction of the ensuing constant electric field in each Fock state. Applying this approach to relativistic $\qu\qb$ and $\qu\qu\qu$ states in QCD results in a bound state equation which provides a reasonable description of the spectrum, including linear Regge trajectories. The equal-time wave functions have unique Lorentz transformation properties, which ensure the correct dependence of the bound state energy on the center-of-mass momentum. The $\qu\qu\qu$ potential is gauge covariant and confines the three quarks in a symmetric way.}

\begin{document}

\section{Introduction} \label{intro}

The perturbative expansion is a powerful tool in the analysis of scattering amplitudes for elementary fields (leptons, quarks and gauge bosons). Bound states (atoms and hadrons) appear as poles in scattering amplitudes which arise from the {\em divergence} of the perturbation series. The study of bound states and their interactions thus requires summing an infinite set of Feynman diagrams. This may be done in the Bethe-Salpeter framework \cite{Salpeter:1951sz}. However, the choice of diagrams to be included in a first approximation, as well as the systematic ordering of the inclusion of the remaining contributions, allows a wide range of approaches, particularly for relativistic bound states.  

Here I shall argue that an $\hbar$ (or equivalently, loop) expansion may serve as a guiding principle in bound state calculations. In the $\hbar \to 0$ limit the phase $\exp(iS/\hbar)$ in the path integral oscillates rapidly, causing the (bosonic) fields to approach their classical values, corresponding to a stationary action $S$. In the absence of field fluctuations there are no loop integrals -- thus the Feynman diagrams of lowest order in $\hbar$ are Born terms. Since each loop brings a factor $\hbar g^2$ the perturbative and loop expansions are equivalent for a given Green function\footnote{The equivalence between the number of loops and powers of $\hbar$ is actually non-trivial, see \cite{Holstein:2004dn}.}. Bound state wave functions on the other hand contain all powers of $g$, yet may 
have no loop contributions, thus decoupling the $\hbar$ and $g$ expansions. This is illustrated by the way in which the Schr\"odinger and Dirac equations emerge from QED perturbation theory.

The Schr\"odinger equation for non-relativistic (NR) atoms like muonium ($e^-\mu^+$) is generated by the sum of ladder diagrams shown in \fig{sladder}. 
%
\EPSFIGURE[h]{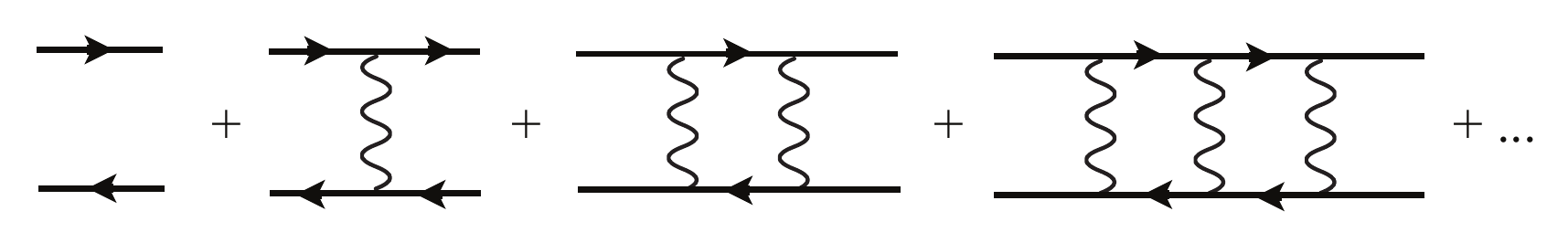,width=.8\columnwidth}{Sum of ladder diagrams in $e^-\mu^+$ scattering which generate a non-relativistic bound state pole.
\label{sladder}}
%
The factor $\alpha=e^2/4\pi$ suppression of each photon exchange is compensated when the propagators in the corresponding loop integral are nearly on-shell, which brings an inverse factor of the relative velocity $v$ between the fermions. Thus for $v \sim \alpha$ all diagrams are effectively of the same order, causing the series to diverge regardless of how small $\alpha$ is. Due to the NR motion Coulomb photon exchange dominates and transfers only 3-momentum. Energy transfer is suppressed by $\alpha$, quenching the loop integrals in \fig{sladder} at leading order. Hence the dynamics reduces to scattering from a fixed $\alpha/r$ Coulomb potential and is described by the Schr\"odinger equation. The repeated scattering from the potential brings powers of $\alpha$ without loops.

%
\EPSFIGURE[h]{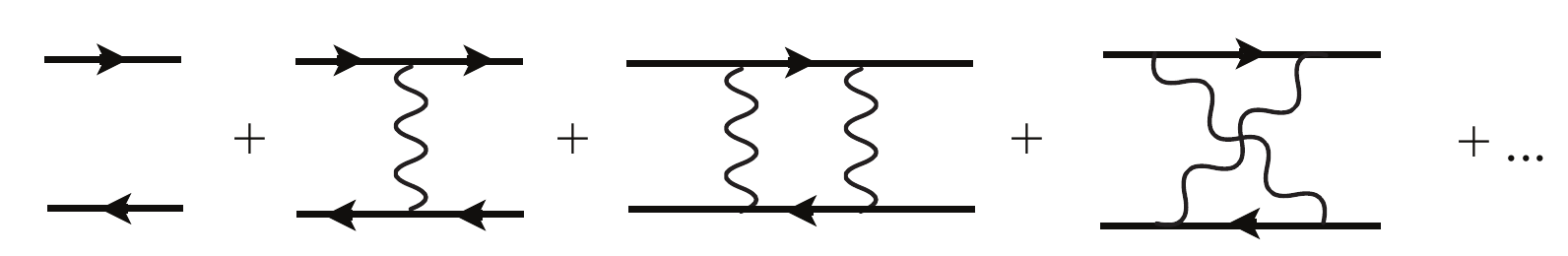,width=.8\columnwidth}{Sum of all uncrossed and crossed photon exchange diagrams which generate Dirac bound states in the limit where one of the fermion masses is large.
\label{dladder}}
%
The Dirac equation for a relativistic electron may be derived in a similar way \cite{Brodsky:1971sk}, by including all crossed photon exchanges as indicated in \fig{dladder}. In the limit of a large muon mass the full sum of crossed and uncrossed photon exchange diagrams reduces at each order in $\alpha$ to the scattering of the electron in a fixed $\alpha/r$ Coulomb potential described by the Dirac equation\footnote{In a Bethe-Salpeter approach the crossed photon exchanges require kernels of unlimited orders in $\alpha$.}. Hence the leading contribution of the sum of multiloop diagrams in \fig{dladder} to the bound state poles  arises from classical Coulomb scattering and in this sense is of lowest order in $\hbar$.

The symmetries of the lagrangian are preserved at each order in $\hbar$. Thus boost invariance requires that the bound state energy $E$ must be related to its CM momentum $\kv$ as $E=\sqrt{M^2+\kv^2}$. Wave functions defined at equal time in different frames must also be related, but that relation is dynamical since the concept of equal time is frame dependent. I find an example of this in the present approach.

Scattering amplitudes at the Born level are insensitive to the $\ieps$ prescription of the propagators. The relative magnitudes of the time-ordered contributions to these amplitudes on the other hand do depend on the prescription. Bound states evaluated at lowest order in $\hbar$ are analogous to Born terms. I show that the energies of Dirac bound states indeed are independent of the $\ieps$ prescription, whereas the equal-time wave functions do depend on the prescription. The standard single-electron Dirac wave function corresponds to using a retarded electron propagator, which only allows forward propagation in time. The use of a Feynman propagator would give Fock states having arbitrarily many electron-positron pairs, arising from backward propagation of the electron in time ($Z$-graphs). The use of retarded propagators thus essentially simplifies bound state analyses at lowest order in $\hbar$.

Gauge theory lagrangians have no time derivative of $A^0$ (unless this is introduced via the gauge fixing). In the $\hbar \to 0$ limit $A^0$ is determined at each instant of time by the positions of the charged constituents. This implies that Fock states with constituents at different positions have distinct $A^0$ fields. The field equations allow a linear $A^0$ potential as a homogeneous (non-perturbative) solution. For the action to be stationary the orientation of the potential needs to be correlated with the locations of the charges in each Fock component. Consequently a linear potential is allowed in a self-consistent semi-classical treatment, maintaining Poincar\'e invariance.

In the next section I outline the main results, which are then discussed in more detail in the following sections.

\section{Outline}

\subsection{QED atoms and the linear potential option} \label{linintro}

I use a hamiltonian approach to field theory bound states in the semi-classical limit, \ie, at lowest order in $\hbar$ (no loops). I first illustrate the procedure for ordinary QED atoms (muonium, $e^-\mu^+$) in their rest frame, whose binding is known to be dominated by the $A^0$ Coulomb potential. This example reminds us that the instantaneous $A^0$ potential determined by the equations of motion is different for each Fock state component. It also clarifies the distinction between the $A^0$ field for a given Fock state and the $A^0$ field that would be measured by an external probe, the latter being given by a superposition of all contributing Fock states.

The determination of $A^0$ from the QED field equation for the Fock component where the electron is located at $\xv_{1}$ and the muon at $\xv_{2}$,
\beq\label{a0eom0}
-\nv_{\xv}^2\, A^0(\xv;\xv_{1},\xv_{2}) = e\delta^3(\xv-\xv_{1})-e\delta^3(\xv-\xv_{2})
\eeq
is unique only up to a homogenous solution. In particular, a contribution linear in $\xv$ may be added,
\beq\label{a0sol2}
A^0(\xv;\xv_{1},\xv_{2}) = \la^2\,\chat\cdot\xv+\frac{e}{4\pi}\left(\inv{|\xv-\xv_{1}|}-\inv{|\xv-\xv_{2}|}\right) \equiv A^0_{lin}+A^0_{coul}
\eeq
where the magnitude $\Lambda$ and the direction $\chat$ of the linear term are free parameters. The gauge field \eq{a0sol2} gives an instantaneous contribution to the action,
\beqa\label{lingauge}
-\inv{4}\int d^3\xv F_{\mu\nu}F^{\mu\nu} = \inv{2}\int d^3\xv\, (\nv A^0)^2 = \inv{2}\la^4\int d^3\xv + \inv{3} e\Lambda^2 \chat\cdot (\xv_{1}-\xv_{2}) -\frac{e^2}{4\pi}\inv{|\xv_{1}-\xv_{2}|}\nn\\
\eeqa
Here I suppressed the infinite self-energy contribution of the Coulomb potential, which is independent of $\xv_{1}$ and $\xv_{2}$ and thus irrelevant for bound state evolution. The first term in \eq{lingauge} is likewise infinite and irrelevant provided $\Lambda$ is a universal constant, $\Lambda \neq \Lambda(\xv_{1},\xv_{2})$. The interference term between the linear and Coulomb contributions is finite and may be evaluated by taking the integration region to be a sphere whose radius $R$ tends to infinity. Using $\nv^2 A^0_{lin}=0$ and Stokes' theorem,
\beq\label{actqed}
\int d^3\xv\,\nv\cdot [(\nv A^0_{lin}) A^0_{coul}]= \lim_{R\to\infty} \int_{|\xv|=R}d{\bs S}\cdot(\nv A^0_{lin}) A^0_{coul}=\inv{3}e\Lambda^2\,\chat\cdot(\xv_{1}-\xv_{2})
\eeq
where terms of \order{1/R^3} in the Coulomb potential could be neglected,
\beq
{\inv{|\xv-\xv_{i}|}}_{|\xv|=R}=\inv{R}\left[1+\frac{\xv\cdot\xv_{i}}{R^2}\right]_{|\xv|=R} + \morder{\inv{R^3}}
\eeq

Stationarity of the action \eq{lingauge} \wrt variations in the direction of the linear potential requires $\chat$ to be parallel to $\xv_{1}-\xv_{2}$. Choosing the sign to give an attractive (rather than repulsive) potential leads to
\beq\label{a0sol3}
A^0(\xv;\xv_{1},\xv_{2}) = \la^2\,\frac{\xv_{1}-\xv_{2}}{|\xv_{1}-\xv_{2}|}\cdot\xv+\frac{e}{4\pi}\left(\inv{|\xv-\xv_{1}|}-\inv{|\xv-\xv_{2}|}\right)
\eeq
with $\la$ a universal constant.
The linear potential allows to incorporate non-perturbative features of the theory at the semi-classical level. The fermion interaction with the $A^0$ field gives
\beq\label{a0int}
e\left[A^0(\xv_{1};\xv_{1},\xv_{2})-A^0(\xv_{2};\xv_{1},\xv_{2})\right]=
e\la^2\,|\xv_{1}-\xv_{2}| - 2\,\frac{e^2}{4\pi}\inv{|\xv_{1}-\xv_{2}|}
\eeq
where I again omitted the infinite, $\xv_{1}$- and $\xv_{2}$-independent Coulomb self-interactions. The Coulomb potential is counted twice since both the electron and the muon interact. However, according to \eq{lingauge} the phase originating from the gauge part of the action $\exp(-\frac{i}{4}\int dt \int d^3\xv F_{\mu\nu}F^{\mu\nu})$, which describes the energy stored in the Coulomb field, reduces the Coulomb potential to its physical value, $V_{C}=-\alpha/r$.

The experimental fact that QED does not confine electric charge compels us to choose $\la_{QED}=0$ and thus recover the standard description of muonium bound only by the Coulomb potential. In Sections \ref{meson} and \ref{baryon} I find corresponding linear potential solutions of the QCD equations of motion for baryons and mesons. Choosing $\la_{QCD} \neq 0$ amounts to a novel boundary condition for perturbation theory, motivated by the confinement of color in QCD.

A linear potential extending to infinity would be unacceptable on physical grounds. However, $A^0(\xv;\xv_{1},\xv_{2})$ of \eq{a0sol3} is the field of only a single Fock state. The field that would be measured by an external probe is a superposition of the contributions from all Fock components. As seen from \eq{a0sol3}, Fock components with opposite separations $\xv_{1}-\xv_{2}$ give opposite contributions to $A^0$ which cancel if the corresponding Fock probabilities are the same. Hence the electric field vanishes outside the bound state, where all Fock states contribute coherently. 

The linear potential contributes at \order{e} since $\la$ is non-perturbative, see \eq{a0int}. Hence in a perturbative sense it dominates the single (Coulomb and transverse) photon exchanges which are of \order{e^2}. This allows the analog of a ``Born term'' for gauge theory bound states, which is exact at lowest order in $\hbar$ (no loops) and at first order in the coupling. Lorentz invariance must then be preserved. In particular, the bound state energy $E$ should have the correct dependence, $E=\sqrt{M^2+\kv^2}$, on the CM momentum $\kv$. The bound state equation that we derive here for wave functions at equal time in all frames in fact has this non-trivial and unique property. By contrast, in QED atoms the $A^0$ Coulomb potential dominates transverse photon exchange {\it only in the rest frame}. The boosted muonium atom gets a leading \order{e^2} contribution from $\ket{e\mu\gamma}$ Fock states with a transverse photon which must be included to get the correct $\kv$ dependence \cite{Jarvinen:2004pi}.

\subsection{Dirac equation in QED -- retarded boundary conditions} \label{diracintro}

Particle production can no longer be neglected when the dynamics is relativistic. An $\ket{e\mu}$ state at $t=0$ will thus with time develop into states with additional electron and muon pairs if the interaction potential $A^0$ is commensurate with the particle masses. Relativistic bound states (defined at equal time of the constituents) therefore have Fock states with arbitrarily many particles. 

The Dirac wave function $\psi(\xv)$ of an electron bound in an external potential is relativistic yet describes the spatial distribution of a single (positive or negative energy) electron. Understanding the apparent absence of multi-particle Fock states in the Dirac wave function turns out to be useful for formulating a relativistic hamiltonian description of $e^-\mu^+$ atoms and hadrons. 

The Green function $G(p^0,\pv)$ of an electron in a static (time-independent) $A^0$ potential (\fig{DiracBS}) satisfies
\beq
G(p^0,\pv) = S + SKG
\eeq
where $S$ is the electron propagator and $K$ the kernel for a single $A^0$ interaction.
%
\EPSFIGURE[h]{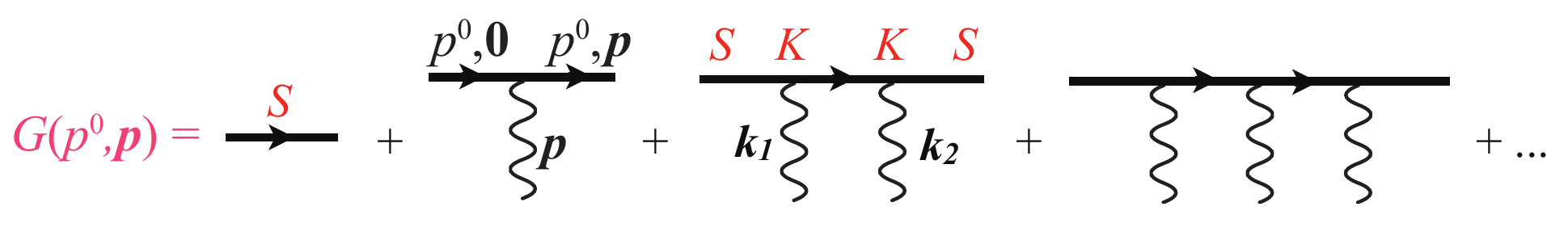,width=1.\columnwidth}{Electron scattering from a static external potential. The energy $p^0$ of the electron does not change during the scattering. The initial and final electron momenta are denoted $(p^0,\bs{0})$ and $p=(p^0,\pv)$.
\label{DiracBS}}
%
The $p^0$ component of the electron's 4-momentum is conserved during the scattering since the static source only transfers 3-momentum. I take the initial and final 3-momenta of the electron to be $\bs{0}$ and $\pv$, respectively.

The Green function has poles at the bound state energies $E_{R}$,
\beq\label{dirpole}
G(p^0,\pv) = \frac{R(E_{R},\pv)}{p^0-E_{R}} + \ldots
\eeq
The pole residue $R(E_{R},\pv)$ satisfies $R = SKR$, or explicitly (with $p^0=E_{R}$)
\beq
R(E_{R},\pv) = \frac{i}{\slashed{p}-m+\ieps}\int\frac{d^3\kv}{\pit}(-ie)\gamma^0 A^0(\kv)\,R(E_{R},\pv-\kv)
\eeq
Multiplying by $\slashed{p}-m$ and Fourier transforming to $(p^0,\xv)$-space we find that the residue $R$ satisfies the Dirac equation,
\beq\label{direq}
\Big[-i\nv\cdot\gv + e\gamma^0 A^0(\xv)+m\Big]R(E_{R},\xv) = E_{R}\gamma^0\,R(E_{R},\xv)
\eeq

In order to display the equal-time Fock states of the bound state given by the Dirac equation we need to time-order the interactions. At \order{e^2} the first (second) diagram on the \rhs of \fig{DiracBS-T} corresponds to the intermediate electron having positive (negative) energy, $E_{i}=\pm \sqrt{\kv_{i}^2+m_{e}^2}$. According to the Feynman $\ieps$ prescription this electron propagates forward (backward) in time, corresponding to an intermediate $\ket{e^-}$ ($\ket{e^-e^+e^-}$) Fock state. At higher orders in $e$ further time orderings contribute. Consequently the bound state has Fock components with arbitrarily many $e^+e^-$ pairs. The creation and destruction of the various Fock components balance to create a bound state which is stationary in time.

\EPSFIGURE[h]{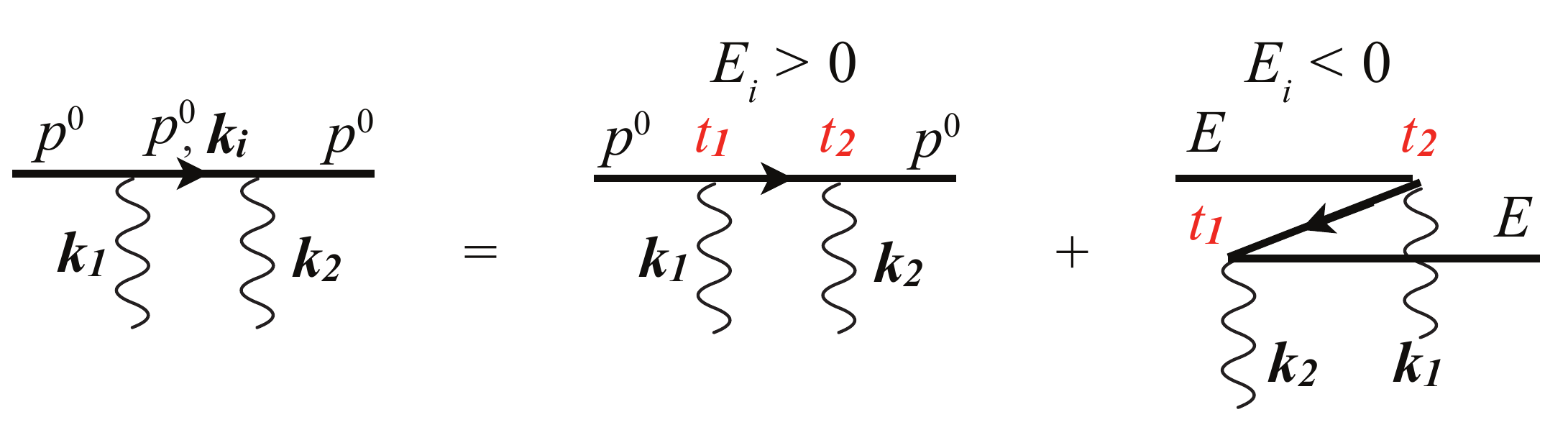,width=1.\columnwidth}{Electron scattering from a static $A^0$ potential at \order{e^2}. The covariant ($p^0$-conserving) diagram on the \lhs splits under time-ordering into the two diagrams on the \rhs The intermediate electron has 3-momentum $\kv_{i}$ and energy $E_{i}=\pm \sqrt{\kv_{i}^2+m_{e}^2}$. The negative energy electron propagates backward in time, implying the creation (annihilation) of an $e^+e^-$ pair at $t_{1}\ (t_{2})$. 
\label{DiracBS-T}}

The Dirac wave function $\psi(\xv)$ describes only a single electron, not the details of the multiparticle bound state dynamics. As seen above, the bound state energies $p^0=E_{R}$ given by the poles of the Green function \eq{dirpole} nevertheless are eigenvalues of the (time independent) Dirac equation \eq{direq} satisfied by the pole residues $R(E_{R},\xv)$. This single particle Dirac wave function may be given the following hamiltonian interpretation.

The static potential conserves the energy component $p^0 >0$ of the electron momentum. Hence the covariant Feynman diagrams in \fig{DiracBS} which build the Green function $G(p^0,\pv)$ do not depend on the Feynman $\ieps$ prescription at the negative energy pole of the electron propagator, $p^0=-\sqrt{\kv_{i}^2+m_{e}^2}+\ieps$. In particular, the bound state energies $E_{R}$ of the Green function \eq{dirpole} will not change if instead of the Feynman propagator of the electron we use the retarded one,
\beq\label{sr}
S_{R}(p^0,\pv) = i\frac{\slashed{p}+m_{e}}{(p^0-E_{p}+\ieps)(p^0+E_{p}+\ieps)}
\eeq
where $E_{p}=\sqrt{\pv^2+m_{e}^2}$. However, this changes the time development of the scattering so that the electron always propagates forward in time,
\beq\label{srt}
S_{R}(t,\pv) = \frac{\theta(t)}{2E_{p}}\left[(E_{p}\gamma^0-\pv\cdot\gv+m_{e})e^{-iE_{p}t} +(E_{p}\gamma^0+\pv\cdot\gv-m_{e})e^{iE_{p}t}\right]
\eeq
Consequently the second ($Z$-)diagram on the \rhs of \fig{DiracBS-T} is absent, while the first diagram refers to the (positive or negative energy) electron moving forward in time. There is only a single electron Fock state and the Dirac wave function describes the distribution of this electron. This single particle picture of the time development obtained using retarded boundary conditions does not correspond to the physical Fock state structure. Furthermore, the argument that the bound state energies ($E_{R}$ in \eq{dirpole}) are independent of the boundary condition only holds in the absence of loops, \ie, in the $\hbar\to 0$ limit\footnote{Loops with retarded propagators vanish since all poles are on the same side of the energy axis.}. 

In the absence of backward propagation the electron becomes localized in space as the propagation time vanishes,
\beq\label{sr0}
\lim_{t\to 0^+}S_{R}(t,\xv) = \gamma^0\delta^3(\xv)
\eeq
which makes a hamiltonian approach feasible. In the operator formalism the retarded boundary condition can be implemented using the ``retarded vacuum'',
\beq\label{retvac}
\ket{0}_{R} = N^{-1}\prod_{\pv,\lambda} d_{\pv,\lambda}^\dag \ket{0}
\eeq
where the product is over all momenta $\pv$ and helicities $\lambda$. The (infinite) normalization factor $N$ is fixed by ${_{R}\bra{0}}0\rangle_{R} = 1$.
The retarded vacuum satisfies
\beq\label{annret}
b_{\pv,\lambda}\ket{0}_{R}=d_{\pv,\lambda}^\dag\ket{0}_{R}=0\hspace{.5cm} {\rm and\ hence\ \ } \psi(x)\ket{0}_{R}=0
\eeq
where $\psi(x)$ is the free (interaction picture) fermion field. Consequently the retarded propagator \eq{srt} is given by the standard operator matrix element in the retarded vacuum,
\beq\label{sr2}
S_{R}(x-y)= {_{R}\bra{0}}\,T[\psi(x)\bar\psi(y)]\,\ket{0}_{R}
\eeq
The negative energy contribution to the propagator arises from the $d^\dag d$ term, which represents the removal of a positive energy antifermion from $\ket{0}_{R}$. The interaction hamiltonian annihilates the retarded vacuum,
\beq\label{intham0}
H_{I}(t)\ket{0}_{R} = e\int d^3\xv\,A^0(\xv)\, \psi^\dag(t,\xv)\psi(t,\xv)\ket{0}_{R} =0
\eeq
which ensures the absence of particle production. In Section \ref{Dirac} I derive the Dirac equation using this hamiltonian approach.

\subsection{Relativistic QCD bound states} \label{ffintro}

The instantaneity of the $A^0$ potential is a consequence of the absence of a $\partial_{0} A^0$ term in the lagrangian. The dominance of $A^0$ over the transverse $\bs{A}$ gauge field components in a perturbative expansion holds for non-relativistic motion (atoms in their rest frame) and more generally with a non-perturbative linear $A^0$ potential as in \eq{a0int}. At lowest order in $\hbar$ the use of the retarded vacuum \eq{retvac} as boundary condition maintains bound state energies while allowing a valence wave function, as discussed above for the Dirac equation.

In Section \ref{meson} I find a QCD meson solution using a hamiltonian approach in the retarded vacuum. In a specific gauge the linear potential \eq{a0sol3} is present only in the commuting gauge fields ($A^0_{3}$ and $A^0_{8}$) of SU$_{3}$. The equation is expected to be exact at leading orders in $\hbar$ and the coupling $g$, giving the equivalent of a Born term for bound states. The $\qu\qb$ color singlet wave function $\chi_{\kv}^{\alpha\beta}(\xv)$, where $\kv$ is the CM momentum of the bound state, $\alpha$ and $\beta$ are Dirac indices and $\xv$ is the spatial separation of the quark pair at an instant of time, satisfies the bound state equation
\beq\label{bse}
-i\nv\cdot\com{\bs{\alpha}}{\chi_{\kv}(\xv)}+\halft\kv\cdot\acom{\bs{\alpha}}{\chi_{\kv}(\xv)}+m_{1}\gamma^0\chi_{\kv}(\xv)-\chi_{\kv}(\xv)\gamma^0 m_{2} = (E-V)\chi_{\kv}(\xv)
\eeq
Here $m_{1}\ (m_{2})$ is the mass of the $\qu\ (\qb)$ constituent, $\bs{\alpha}=\gamma^0\gv$ is a standard Dirac matrix and $V(\xv)=g\la^2|\xv|$ is the linear potential. This equation is similar in form to the one proposed by Breit \cite{Breit:1929zz} already in 1929. Here the potential is purely linear since the derivation is valid only at leading order in the gauge coupling $g$. The properties of this equation (with $\kv=0$ and $m_{1}=m_{2}$) was studied phenomenologically \cite{Geffen:1977bh} for a linear + Coulomb potential (see also \cite{Krolikowski:1992fy} and references therein). It was also derived using a stationary phase approximation with retarded boundary conditions \cite{Hoyer:1983xh}.

Due to the underlying gauge invariance it is perhaps not surprising that the energy $E$ of the bound state appears in the canonical form $E-eA^0$ in \eq{bse}, while the CM momentum $\kv$ is kinematical since $\bs{A}=0$. The radial wave functions have singularities at $r=0$ and at $E-V(r)=0$. Requiring local normalizability at these points gave quantized energy levels and a phenomenologically relevant spectrum, including asymptotically linear Regge trajectories \cite{Geffen:1977bh}. The bound state equation reduces to the Schr\"odinger equation in the non-relativistic limit ($V \ll E$), in which case the singular point $E-V(r)=0$ moves to $r=\infty$.

At large quark separations $r$ the linear potential $V(r)$ dominates the \rhs of \eq{bse}, and can be balanced only by a large derivative term on the \lhs Hence the $\qu\qb$ wave function is rapidly oscillating at large $r$, $\chi\sim\exp(ig\la^2 r^2/4)$, and has an $r$-independent probability density. Since this wave function was derived using retarded boundary conditions it indirectly (via its negative energy components) describes the virtual ``sea'' quarks obtained with physical boundary conditions. The normalization $\propto r$ of the wave function \eq{bse} plausibly reflects the quark pairs created by the large potential energy $V(r) \propto r$. This question deserves further study.

Since the wave function $\chi_{\kv}(\xv)$ describes constituents at equal time for all CM momenta $\kv$ it is not explicitly Lorentz covariant. Nevertheless, if the bound state equation \eq{bse} is accurate to lowest order in $\hbar$ and $g$ the bound state energy must have the correct dependence on the CM momentum, $E(\kv) = \sqrt{\kv^2+M^2}$. Remarkably, this turns out to be the case \cite{Hoyer:1985tz}. The $\kv$-dependence of the wave function is explicit in 1+1 dimensions. In 3+1 dimensions $\chi_{\kv}(\xv)$ with $\xv \parallel \kv$ can be expressed in terms of the $\kv=0$ wave function. This also holds for the first derivative of $\chi_{\kv}$ \wrt $\xv \perp \kv$, which (numerically) allows to determine $\chi_{\kv}(\xv)$ for all $\xv$ from the $\kv=0$ solution. 

The $\kv$-dependence of the wave function $\chi_{\kv}$ and of its energy eigenvalues $E(\kv)$ are found as follows \cite{Hoyer:1985tz}. For $\kv=(0,0,k)$ along the $z$-axis, the bound state equation \eq{bse} for $\chi_{\kv}(0,0,z)$ and its first derivatives $\partial_{i}\chi_{\kv}(0,0,z),\ (i=x,y)$ become independent of $k$ when the coordinate $z$ is expressed in terms of the ``boost invariant'' variable $s$ defined by
\beq\label{sdef}
dz=\frac{2}{E(k)-V(z)}ds
\eeq
In the non-relativistic regime, $V\ll E$, this is just the Lorentz contraction expected for distances measured at equal time in every frame. The fact that the transformation is governed by the canonical energy $E-V(z)$ means that the contraction rate for a relativistic wave function is $z$-dependent. The Lorentz covariance of \eq{bse} requires furthermore that the square of the canonical four-momentum $P=(E-V,0,0,k)$ is frame independent when expressed as a function of $s$. For a linear potential $V(z)=g\la^2 |z|$ the relation \eq{sdef} integrates (taking $z>0$) to
\beq
s=\halft z(E-\halft V)
\eeq
With $E^2-k^2=M^2$ we then have indeed that
\beq
P^2 = (E-V)^2-k^2 = M^2-4g\la^2 s
\eeq
is independent of $k$. This seeming coincidence only holds for a linear potential.

For baryons I find a corresponding $\qu\qu\qu$ bound state equation and a gauge where the wave function is fully antisymmetric in color. Stationarity of the action implies that the color electric fields $\nv A^0_{3}$ and $\nv A^0_{8}$ point in separate directions, thus confining all three quarks. The relative strength $A^0_{3}/A^0_{8}$ depends on the positions $\xv_{1},\xv_{2},\xv_{3}$ of the quarks in such a way that the potential is the same for all permutations of quark colors. This allows (in a specific gauge) the simple color structure $\epsilon_{ABC}$ of the wave function, as assumed in the quark model.

\section{Non-Relativistic QED Bound States} \label{nr}

Some aspects of the methods sketched above may be illustrated in the familiar case of non-relativistic (NR) QED atoms, specifically muonium $(e^-\mu^+)$. I use an equal-time, hamiltonian field theory approach which is close in spirit to the usual treatment in quantum mechanics. However, rather than reducing the problem to relative motion in a fixed central potential I determine the instantaneous $A^0$ field by taking matrix elements of the QED operator equations of motion, 
\beq\label{amueom}
\partial_{\mu} F^{\mu\nu}(x)-e\sum_{i=e,\mu}\bar\psi_{i}(x)\gamma^\nu \psi_{i}(x)=0
\hspace{1cm} {\rm (EOM)}
\eeq
The resulting $A^0$ depends on the matrix element, and specifically on the positions of the electron $(\xv_{1})$ and muon $(\xv_{2})$. Hence each Fock component of the bound state is assigned a distinct potential $A^0(\xv;\xv_{1},\xv_{2})$. This is made possible by the instantaneity of $A^0$, which in turn is a consequence of the absence of a time derivative $\partial_{0}A^0$ in gauge theory lagrangians. The Coulomb gauge condition ($\nv\cdot \av=0$) maintains the instantaneity of $A^0$ and in this sense is preferable to covariant gauge fixing ($\partial_{\mu}A^\mu=0$).

In the NR limit we may ignore
fermion pair production (I address the relativistic effects in Section \ref{Dirac}). The dominance of the $A^0$ potential\footnote{This holds only in the atomic rest frame. For atoms in relativistic CM motion also $\ket{f\bar f\gamma}$ Fock states with a transverse photon contribute at lowest order \cite{Jarvinen:2004pi}.} and its instantaneous nature then implies that only two-body $\ket{e^-\mu^+}$ Fock states contribute at lowest \order{e^2} in the coupling.

I parametrize the ${e^-\mu^+}$ bound state at time $t=0$ as
\beq\label{ffstate}
\ket{E,t=0}=\int d^3\yv_{1}d^3\yv_{2}\,\psi_{e}^\dag(t=0,\yv_{1})\chi(\yv_{1},\yv_{2})\psi_{\mu}(t=0,\yv_{2})\ket{0}
\eeq
where the wave function $\chi(\yv_{1},\yv_{2})$ is a $4\times 4$ matrix in Dirac space\footnote{I use a relativistic notation here and later reduce the wave function to its leading NR components.}. The free fermion operators of the Interaction Picture are as usual
\beq\label{ffield}
\psi(x) = \int\dpve\sum_{\lambda}\left[u(\pv,\lambda)e^{-ip\cdot x}b_{\pv,\lambda} + v(\pv,\lambda)e^{ip\cdot x}d_{\pv,\lambda}^\dag \right]
\eeq
and satisfy the canonical anticommutation relation 
\beq\label{canon}
\acom{\psi_{\alpha}(t,\xv)}{\psi_{\beta}^\dag(t,\xv')}=\delta^3(\xv-\xv') \delta_{\alpha\beta}
\eeq

The matrix elements of the EOM \eq{amueom} for each $\ket{e^-\mu^+}$ Fock component of the bound state \eq{ffstate} at $t=0$ should vanish, 
\beq\label{eomatr}
\bra{0}\psi_{\mu\beta}^\dag(0,\xv_{2}) \psi_{e\alpha}(0,\xv_{1})\,{\rm (EOM)}\, \ket{E,0}=0
\eeq
Since the Fock states do not contain physical (transverse) photons only the classical (instantaneous) $A^0$ field survives in $F^{\mu\nu}(x)$ at lowest order in the coupling $e$. For $\nu=0$ in \eq{amueom} the constraint \eq{eomatr} is
\beqa\label{a0eom}
 \chi_{\alpha\beta}(\xv_{1},\xv_{2})\left[-\nv^2 A^0(\xv)\right] &=& e\sum_{i=e,\mu}\bra{0}\psi_{\mu\beta}^\dag(0,\xv_{2}) \psi_{e\alpha}(0,\xv_{1})\,\psi_{i}^{\dag}(0,\xv) \psi_{i}(0,\xv)\, \ket{E,0}\nn\\
&=& e\left[\delta^3(\xv-\xv_{1})-\delta^3(\xv-\xv_{2})\right]\chi_{\alpha\beta}(\xv_{1},\xv_{2})
\eeqa
where I used (neglecting multi-particle states in the NR limit, hence $\psi_{e}(x)\ket{0}\to 0$)
\beq\label{frel}
\psi_{e}(0,\xv)^{\dag} \psi_{e}(0,\xv)\ket{E,0} = \psi_{e}^{\dag}(0,\xv) \int d^3\yv_{2}\chi(\xv,\yv_{2}) \psi_{\mu}(0,\yv_{2})\ket{0}
\eeq
The standard solution is
\beq\label{a0sol1}
A^0(\xv;\xv_{1},\xv_{2}) = \frac{e}{4\pi}\left(\inv{|\xv-\xv_{1}|}-\inv{|\xv-\xv_{2}|}\right)
\eeq
The interpretation of this result differs from the case where $A^0(\xv)$ is regarded as a {\it fixed external Coulomb potential} centered at $\xv$, which is sampled by the charged constituents according to their positions $\xv_{1}$ and $\xv_{2}$. Now there is no external field but rather a bound state \eq{ffstate} which is a superposition of Fock states. The gauge field $A^0$ is {\it constrained for each Fock component} and each instant of time by the QED equation of motion. At large $|\xv|$ the Coulomb field \eq{a0sol1} vanishes faster than $1/|\xv|$ as appropriate for a neutral state. Furthermore,  \eq{a0sol1} is the field of a single Fock component. A measurement of $A^0$ by an external probe would involve a sum over all Fock contributions weighted by their probabilities. This ensures, \eg, the cancellation of the dipole field outside $S$-wave bound states.

For $\nu=j\ (=1,2,3)$ in the EOM \eq{amueom} the $A^0$ field does not contribute in the matrix element \eq{eomatr},
\beq\label{ajeom}
 \chi_{\alpha\beta}(\xv_{1},\xv_{2})\cdot 0 =e\,\delta^3(\xv-\xv_{1})(\gamma^0\gamma^{j}\chi)_{\alpha\beta}-e\,\delta^3(\xv-\xv_{2})(\chi\gamma^0\gamma^{j})_{\alpha\beta} =0
\eeq
This is satisfied at lowest order in the coupling $e$ for NR QED atoms since the space components $\bs{j}$ of the constituent currents are suppressed compared to their charge densities $j^0$ by a factor $p/m\sim \alpha$, where $p$ is the relative momentum and $m$ the mass scale of the constituents. In analogy with the scaling of the upper and lower components of a Dirac wave function I expect (and verify below) that the wave function $\chi(\xv_{1},\xv_{2})$  has the Dirac structure, in $2\times 2$ block form,
\beq\label{chisize}
\chi =\left(\begin{array}{cc}
  \chi_{11} & \chi_{12} \\ \chi_{21} & \chi_{22}
 \end{array}\right) \ ,\hspace{1cm} 
 \left\{\begin{array}{ccc}
  \chi_{12} &=& \morder{1} \\  \chi_{11},\ \chi_{22} &=& \morder{p/m} \\ 
  \chi_{21} &=& \morder{p^2/m^2}
 \end{array}\right.
\eeq
For the dominant Fock components described by $\chi_{12}$ the \rhs of the EOM \eq{ajeom} is thus suppressed by $p/m$ due to the off-diagonal nature of $\gamma^0\gamma^j$. The EOM \eq{a0eom} for the other components $\chi_{ij}$ is of higher order in $e^2$ due to the suppression of the corresponding wave functions.

Having determined the $A^0$ field \eq{a0sol1} for each Fock component and verified that the EOM is consistent with $\av=0$ we may impose stationarity in time on the Fock amplitudes,
\beq\label{bsdef}
\phi_{\alpha\beta}(t;\xv_{1},\xv_{2}) \equiv \bra{0}\psi_{\mu\beta}^\dag(t,\xv_{2}) \psi_{e\alpha}(t,\xv_{1})\ket{E,t} = e^{-iEt} \phi_{\alpha\beta}(t=0;\xv_{1},\xv_{2})
\eeq
where $\phi_{\alpha\beta}(t=0;\xv_{1},\xv_{2})= \chi_{\alpha\beta}(\xv_{1},\xv_{2})$ follows from the anticommutation relation \eq{canon}.

The time dependence of the fermion fields is explicit in \eq{ffield} while that of the state $\ket{E,t}$ is given by the interaction Hamiltonian 
\beq\label{intham}
H_{I}(t) = e\int d^3\xv\,A^0(\xv)\sum_{f=e,\mu} \psi_{f}^\dag(t,\xv)\psi_{f}(t,\xv)
\eeq
$H_{I}$ actually appears with a factor $\halft$ in Coulomb gauge quantization 
(see, \eg, Appendix A of \cite{sak67}). 
Equivalently, and perhaps more intuitively, we may consider \eq{intham} as giving the interaction energy of the electron and muon with the Coulomb field, but then need to consider also the energy $E_{A}$ stored in the field, since it depends on the Fock state,
\beqa\label{phase}
\exp\left[idt\int d^3\xv(-\quart F_{\mu\nu}F^{\mu\nu})\right] &=& \exp\left[-\frac{idt}{2}  \int d^3\xv\,A^0(\xv)\nv^2A^0(\xv)\right] \nn\\
&=& \exp\left[-idt\,\frac{e^2}{4\pi}  \inv{|\xv_{1}-\xv_{2}|}\right] \equiv \exp(-idt E_{A})
\eeqa
where I discarded an $\xv_{1},\xv_{2}$-independent infinite phase (self-energy). This contribution cancels half of the $e,\mu$ interaction energies arising from \eq{intham}.

The stationarity condition for the bound state wave function \eq{bsdef} is then (at $t=0$),
\beqa\label{ffcond}
i\frac{d\phi_{\alpha\beta}(0;\xv_{1},\xv_{2})}{dt} = \bra{0} i\frac{d\psi_{\mu\beta}^\dag(0,\xv_{2})}{dt} \psi_{e\alpha}(0,\xv_{1})\ket{E,0} &+& i\psi_{\mu\beta}^\dag(0,\xv_{2})\frac{d\psi_{e\alpha}(0,\xv_{1})}{dt}\ket{E,0}\nn\\
 +\bra{0}\psi_{\mu\beta}^\dag(0,\xv_{2}) \psi_{e\alpha}(0,\xv_{1})[H_{I}(0)+E_{A}]\ket{E,0} &=& E\,\phi_{\alpha\beta}(0;\xv_{1},\xv_{2})
\eeqa
Using \eq{frel} and \eq{canon} gives
\beqa\label{hi}
\bra{0}\psi_{\mu\beta}^\dag(0,\xv_{2}) \psi_{e\alpha}(0,\xv_{1})H_{I}(0)\ket{E,0} &=& 
\left[eA^0(\xv_{1})-eA^0(\xv_{2})\right]\chi(\xv_{1},\xv_{2}) \nn\\
 &=& -2\,\frac{e^2}{4\pi} \inv{|\xv_{1}-\xv_{2}|}\chi(\xv_{1},\xv_{2})
\eeqa
where I again discarded the infinite self-energy terms. Due to the electron interacting with the field of the muon and {\it vice versa} the potential energy is twice the physical value. In the bound state equation \eq{ffcond} this is reduced by the field energy $E_{A}$ of \eq{phase} to the standard Coulomb potential
\beq\label{pot1}
V(\xv_{1},\xv_{2}) =  -\frac{e^2}{4\pi} \inv{|\xv_{1}-\xv_{2}|}
\eeq
Altogether the stationarity condition \eq{ffcond} gives,
\beq\label{ffqed}
\gamma^0(-i\nv_{1}\cdot\gv +m_{e})\chi(\xv_{1},\xv_{2}) - \chi(\xv_{1},\xv_{2})\gamma^0(i\nv_{2}\cdot\gv +m_{\mu}) = [E-V(\xv_{1},\xv_{2})] \chi(\xv_{1},\xv_{2})
\eeq
where $\nv_{2}\equiv \partial/\partial\xv_{2}$ operates to the left. This is identical in form with the relativistic bound state equation that I derive in Section \ref{meson}, but presently valid only in the non-relativistic limit where particle production and the transverse components of the gauge field $\av$ may be ignored. 

The reduction of \eq{ffqed} to the non-relativistic Schr\"odinger equation is analogous to the standard reduction of the Dirac equation. In the CM, $\chi(\xv_{1},\xv_{2}) \to \chi(\xv_{1}-\xv_{2})\equiv \chi(\xv)$. Ignoring the \order{p^2/m^2} component $\chi_{21}$ in \eq{chisize} the bound state equation \eq{ffqed} is in component form
\beq\label{ffcomp}
\left(\begin{array}{cc}
i\nv\chi_{12}\cdot\sv & -i\sv\cdot\nv\chi_{22}+i\nv\chi_{11}\cdot\sv \\
-i\sv\cdot\nv\chi_{11}+i\nv\chi_{22}\cdot\sv & -i\sv\cdot\nv\chi_{12}
\end{array}\right) =
\left(\begin{array}{cc}
2m_{\mu}\,\chi_{11}\ & (E_{b}-V)\chi_{12} \\
0 & 2m_{e}\,\chi_{22}
\end{array}\right)
\eeq
where on the \rhs I defined the \order{p^2/m} binding energy $E_{b}$ through $E=m_{e}+m_{\mu}+E_{b}$ and approximated $E-V-m_{e}+m_{\mu}\simeq 2m_{\mu}$, $E-V+m_{e}-m_{\mu}\simeq 2m_{e}$. Substituting
\beq
\chi_{11} = \frac{i}{2m_{\mu}}\nv\chi_{12}\cdot\sv \hspace{2cm}
\chi_{22} = \frac{-i}{2m_{e}}\nv\cdot\sv\chi_{12}
\eeq\label{scheq}
in \eq{ffcomp} gives the Schr\"odinger equation for $\chi_{12}$,
\beq
\left(-\frac{\nv^2}{2m_{e}}-\frac{\nv^2}{2m_{\mu}} +V\right)\chi_{12}=E_{b}\chi_{12}
\eeq
As expected the Schr\"odinger equation is independent of the $2\times 2$ matrix structure of $\chi_{12}$, and hence of the electron and muon spin orientations.

\section{The Dirac Equation} \label{Dirac}

The electron equation of motion,
\beq\label{feom}
(i\slashed{\partial}-e\slashed{A}-m_{e})\psi(x)=0
\eeq
follows directly from the QED lagrangian and has precisely the form of the Dirac equation. However, \eq{feom} is an exact operator relation whereas the Dirac equation gives an approximate c-numbered amplitude $\psi(x)$ for electron scattering in an external gauge field. The Dirac equation may be derived from the field theoretic Bethe-Salpeter equation by considering $e^{-}\mu^+$ scattering in the limit where the muon mass tends to infinity \cite{Brodsky:1971sk}. Somewhat surprisingly it turns out that the interaction kernel cannot be restricted to any finite order in the coupling $e$ but must include an unlimited number of crossed photon exchanges, as also seen from \fig{dladder}. On the other hand, the Dirac equation does not include loop corrections on the electron propagator or vertices.

As already discussed in Section \ref{diracintro}, and pictorially shown in \fig{DiracBS-T}, the interaction hamiltonian \eq{intham} creates fermion pairs in the perturbative vacuum. Nevertheless, the Dirac (c-numbered) wave function $\psi(t,\xv)$ at a given time depends on the coordinate $\xv$ of a single (positive or negative energy) electron. This may be understood by considering electron scattering in the retarded vacuum \eq{retvac}, $\rket\equiv N^{-1}\prod_{\pv,\lambda} d_{\pv,\lambda}^\dag \ket{0}$, where the product extends over all positron momenta $\pv$ and helicities $\lambda$. With retarded boundary conditions both positive and negative energy electrons propagate forward in time, $Z$-diagrams are absent and  according to \eq{intham0} there is no particle production,
\beq\label{hretvac}
H_{I}\ket{0}_{R} = 0
\eeq

In a covariant perturbative description of electron scattering from a static potential $A^0(\xv)$ the electron energy $p^0$ is unchanged by the scattering. Hence, for $p^0>0$ and in the absence of loop integrals, the $\ieps$ prescription at the negative energy pole $p^0 = -\sqrt{\pv^2+m_{e}^2}$ of the electron propagator $S(p^0,\pv)$ is irrelevant. Consequently the bound state energies, \ie, the locations of the poles in the $S$-matrix, are the same for Feynman $\ket{0}$ and retarded $\rket$ boundary conditions. On the other hand, the Fourier transform $p^0 \to t$ is sensitive to the $\ieps$ prescription at the negative energy pole. The Dirac wave function $\psi(t,\xv)$ describes the single electron bound state obtained with retarded boundary conditions, thus hiding its true multi-pair Fock state structure.

It is straightforward to derive the standard Dirac equation for an electron bound by an external potential $A^0(\xv)$ using the hamiltonian method of Section \ref{nr} in the $\rket$ vacuum. The bound state at $t=0$ is parametrized in terms of its 4-component wave function $\varphi(\xv)$ as
\beqa\label{dstate}
\ket{E,t=0} &\equiv& \int d^3\xv \psi^\dag(t=0,\xv)\varphi(\xv)\rket \nn\\
&=& \int\dpve \sum_{\lambda}\left[u^\dag(\pv,\lambda)\varphi(\pv)b_{\pv,\lambda}^\dag\ket{0}_{R}+ v^\dag(-\pv,\lambda)\varphi(\pv)d_{-\pv,\lambda}\ket{0}_{R}\right]
\hspace{1cm}
\eeqa
The negative energy components of $\varphi(\pv)$ describe a state where $d_{-\pv,\lambda}$ has removed a positive energy antifermion from $\ket{0}_{R}$. For the bound state to be stationary in time each Fock state amplitude $\phi(t,\xv)$ must satisfy
\beq
\phi(t,\xv) \equiv \rbra\psi(t,\xv)\ket{E,t} = e^{-iEt}\phi(0,\xv)
\eeq
where $\phi(0,\xv) = \varphi(\xv)$ follows from $\acom{\psi_{\alpha}(t,\xv)}{\psi_{\beta}^\dag(t,\xv')}=\delta^3(\xv-\xv')\,\delta_{\alpha\beta}$ and $\psi(x)\rket=0$.

The time dependence of $\ket{E,t}$ is given by the interaction Hamiltonian \eq{intham}.
The stationarity requirement for the bound state at $t=0$ is then
\beq\label{dbound}
i\frac{d\phi(0,\xv)}{dt} = \rbra i\frac{d\psi(0,\xv)}{dt}\ket{E,0} + \rbra\psi(0,\xv)H_{I}\ket{E,0} = E \phi(0,\xv)
\eeq
The interaction picture fields satisfy
\beq\label{psieom}
i\frac{d\psi(t,\xv)}{dt} = \gamma^0(-i\nv\cdot\gv +m_{e})\psi(t,\xv)
\eeq
and, making use of $\rbra \psi^\dag(t,\xv)=0$,
\beq
\rbra \psi(0,\xv)H_{I}\ket{E,0} = eA^0(\xv)\varphi(\xv)
\eeq
Using these relations in \eq{dbound} gives the Dirac equation for the wave function $\varphi(\xv)$ of a bound state of energy $E$ in the external potential $A^0(\xv)$,
\beq\label{diraceq}
(-i\nv\cdot\gv+e\gamma^0 A^0(\xv) +m_{e})\varphi(\xv) = E\gamma^0 \varphi(\xv)
\eeq

The time development of fermion states may equivalently be derived using functional integral methods \cite{Hoyer:1982ni}. The retarded vacuum $\ket{0}_{R}$ of \eq{retvac} is then seen to correspond to a vacuum without a Dirac sea. I briefly recall this in the Appendix.

\section{Meson bound state equation in QCD} \label{meson}

I now apply the hamiltonian method demonstrated for non-relativistic QED atoms in Section \ref{nr} to relativistic color singlet $u\bar d$ mesons, using the standard QCD lagrangian 
\beqa\label{qcdlag}
\lcal_{QCD} &=& -\quart F_{a}^{\mu\nu}F_{\mu\nu}^{a} + \sum_{f=u,d}\bar\psi_{f}^{A}(i\sla{\partial}-g\sla{A}_{a}T^{a}_{AB}-m_{f})\psi_{f}^{B} \nn\\
F_{a}^{\mu\nu} &=& \partial^\mu A_{a}^\nu - \partial^\nu A_{a}^\mu - gf_{abc}A_{b}^\mu A_{c}^\nu
\eeqa
The explicit description of particle production in the perturbative vacuum is avoided by using the retarded vacuum, shown in Section \ref{Dirac} to give the Dirac equation and correct bound state energies at lowest order in $\hbar$. The filling of all $\bar u$ and $d$ states now includes a product also over colors $A$,
\beq\label{retvac3}
\ket{0}_{R} = N^{-1}\prod_{\pv,\lambda,A} d_{u}^{A\dag}(\pv,\lambda)\, b_{d}^{A\dag}(\pv,\lambda) \ket{0}
\eeq
ensuring that the interaction picture fields satisfy
\beq\label{fann2}
\psi_{u}^{A}(t,\yv)\ket{0}_{R} = \psi_{d}^{A\dag}(t,\yv)\rket =0
\eeq
The filling of all $\bar u$ states may be equivalently expressed in coordinate space,
\beq
\prod_{\pv,\lambda,A} d_{A}^{\dag}(\pv,\lambda)\ket{0} \propto \prod_{\xv,A} \psi_{A}(t,\xv)\ket{0}
\eeq
A local SU(3) gauge transformation $U$ transforms $\psi_{A}(x) \to \sum_{B} U_{AB}(x)\psi_{B}(x)$ and hence
\beq
\psi_{1}\,\psi_{2}\,\psi_{3} \to \psi_{1}\,\psi_{2}\,\psi_{3}\sum_{A,B,C}\veps_{ABC}\,U_{1A}U_{2B}U_{3C} = \psi_{1}\,\psi_{2}\,\psi_{3}\det U =\psi_{1}\,\psi_{2}\,\psi_{3}
\eeq
where the antisymmetric tensor $\veps_{ABC}$ arose from anticommuting the fermion operators. Hence the retarded vacuum \eq{retvac3} is a color singlet.  It is also Lorentz invariant, since $U_{\la}d^{\dag}(\pv,\lambda)\,U_{\la}^{-1}=d^{\dag}(\la\pv,\lambda)$ only amounts to a reordering of the momenta.

The $u\bar d$ bound state at $t=0$ is similarly to \eq{ffstate} expressed as
\beq\label{ffqcd}
\ket{E,t=0}=\int d^3\yv_{1}d^3\yv_{2}\,\psi_{u}^{A\dag}(t=0,\yv_{1})\chi^{AB}(\yv_{1},\yv_{2})\psi_{d}^{B}(t=0,\yv_{2})\ket{0}_{R}
\eeq
The wave function describing the state \eq{ffqcd} is gauge dependent. For the state to be invariant under time-independent gauge transformations $\psi(t,\xv) \to U(\xv)\psi(t,\xv)$ we need
\beq
\chi(\yv_{1},\yv_{2}) \to U(\yv_{1})\chi(\yv_{1},\yv_{2}) U^{\dag}(\yv_{2})
\eeq
The stationarity condition for the bound state wave function corresponding to \eq{bsdef} is
\beq\label{bsqcd}
\phi_{\alpha\beta}^{CD}(t;\xv_{1},\xv_{2}) \equiv \rbra\psi_{d\beta}^{D\dag}(t,\xv_{2}) \psi_{u\alpha}^{C}(t,\xv_{1})\ket{E,t} = e^{-iEt} \phi_{\alpha\beta}^{CD}(t=0;\xv_{1},\xv_{2})
\eeq
where $\phi_{\alpha\beta}(t=0;\xv_{1},\xv_{2})= \chi_{\alpha\beta}(\xv_{1},\xv_{2})$ follows from \eq{fann2} and the anticommutation relation
\beq\label{canon2}
\acom{\psi_{\alpha}^{A}(t,\xv)}{\psi_{\beta}^{B\dag}(t,\xv')}=\delta^3(\xv-\xv')\, \delta_{\alpha\beta} \,\delta^{AB}
\eeq
As in the non-relativistic case \eq{eomatr}, matrix elements of the QCD operator equations of motion should be satisfied to leading order in $g$ for each Fock state,
\beq\label{bseom}
\rbra\psi_{d\beta}^{D\dag}(t,\xv_{2}) \psi_{u\alpha}^{C}(t,\xv_{1}) \Big[\partial_{\mu}F_{a}^{\mu\nu}+gf_{abc}F_{b}^{\mu\nu}A_{\mu}^c - g\sum_{f=u,d}\bar\psi_{f}^{A}\gamma^\nu T_{a}^{AB}\psi_{f}^{B}\Big]\ket{E,t} =0
\eeq
From \eq{fann2} and \eq{ffqcd} we then get the following constraints on the gauge fields at $t=0$:
\beqa\label{bseom2}
\chi^{CD}(\xv_{1},\xv_{2}) \left[\partial_{\mu}F_{a}^{\mu\nu}+gf_{abc}F_{b}^{\mu\nu}A_{\mu}^c\right] &=& g\,\delta^3(\xv-\xv_{1})T_{a}^{CA}\gamma^0\gamma^\nu \chi^{AD}(\xv_{1},\xv_{2})\nn\\
&-& g\,\delta^3(\xv-\xv_{2})\chi^{CA}(\xv_{1},\xv_{2})\gamma^0\gamma^\nu T_{a}^{AD} 
\eeqa
The sources generate fields of \order{g}, whose interactions contribute to bound state evolution at order $g^2$ (\eg, single gluon exchange between the quarks). However, the EOM \eq{bseom2} allows also an \order{g^0} instantaneous field
\beq\label{homsol}
A^0_{a}(\xv) = \la^2_{a}\,\chat_{a}\cdot\xv
\eeq 
as a homogenous solution ($\nv^2 A^0_{a} = 0$) which contributes to bound state evolution at \order{g}. I shall consider only terms of this order, and thus neglect the \order{g^2} contributions from the vector potentials $A_{a}^j$ ($j=1,2,3$). As an {\it ansatz} for a bound state solution I assume the existence of a gauge where the wave function is diagonal in color,
\beq\label{chising}
\chi^{AB}(\yv_{1},\yv_{2}) = \delta^{AB} \chi(\yv_{1},\yv_{2})
\eeq
In order that the quark interactions with the gauge fields do not generate color non-diagonal  Fock states only the commuting $a=3,8$ components of the $A_{a}^0$ gauge fields \eq{homsol} can be non-vanishing in the gauge \eq{chising}. 

The action gets an \order{g} contribution from the interference between the \order{g^0} fields \eq{homsol} and the \order{g} fields $A^0_{a=3,8}$ generated by the sources in the EOM \eq{bseom2}. Due to
$f_{a38}=0$ the gluon self-interaction term does not contribute at lowest order. With $\nu=0$ and quarks of color $C$ the EOM constraint becomes
\beq\label{bseom3}
-\nv^2 A_{a}^{0}(\xv) = g\,T_{a}^{CC}\left[\delta^3(\xv-\xv_{1})
- \delta^3(\xv-\xv_{2})\right]
\eeq
(no sum over the quark color $C$). With the homogeneous solution \eq{homsol} we have then for
$a = 3,8$ the instantaneous potential
\beq\label{a0qcd}
A^0_{a}(\xv;\xv_{1},\xv_{2},C) = \la^2_{a}\,\chat_{a}\cdot\xv+\frac{gT^{CC}_{a}}{4\pi}\left(\inv{|\xv-\xv_{1}|}-\inv{|\xv-\xv_{2}|}\right) \hspace{1cm} (a = 3,8)
\eeq
As in the non-relativistic QED case \eq{a0sol1} the field $A^0_{a}$ depends on the positions of the quarks, and now also on their color. Since \eq{a0qcd} is a solution of the EOM the action is stationary under {\it local} variations of $A_{a}^0$, for any constants $\la_{a}$ and unit vectors $\chat_{a}$. However, variations of these parameters is a {\it global} variation which can affect the action. In fact,
\beqa\label{lingauge2}
-\inv{4}\sum_{a}\int d^3\xv F_{\mu\nu}^{a}F^{\mu\nu}_{a} &=& \inv{2}\sum_{a}\int d^3\xv\, (\nv A_{a}^0)^2 \\
&=&\sum_{a=3,8}\left[\inv{2}\la_{a}^4\int d^3\xv + \inv{3} g\Lambda_{a}^2\, T_{a}^{CC}\chat_{a}\cdot (\xv_{1}-\xv_{2}) +\morder{g^2}\right] \nn
\eeqa
The parameter $\sum_{a=3,8}\la_{a}^4$ is multiplied by the (infinite) volume of space. This term does not affect bound state evolution provided it is the same for all Fock components. Hence 
\beq\label{lamdef}
\la^4 \equiv\sum_{a=3,8}\la_{a}^4 
\eeq
should be a universal constant, independent of $\xv_{1},\xv_{2}$ and the quark color $C$. The \order{g} interference term is finite and was evaluated as in \eq{actqed}. It is stationary \wrt variations of the unit vectors $\chat_{a}$ provided $\chat_{a}\parallel \xv_{1}-\xv_{2}$. Choosing $\chat_{a}=T_{a}^{CC}(\xv_{1}-\xv_{2})/|T_{a}^{CC}(\xv_{1}-\xv_{2})|$ gives (as seen below) an {\it attractive} linear potential $\propto\sum_{a} g\la_{a}^2|T_{a}^{CC}(\xv_{1}-\xv_{2})|$ between quarks of color $C$. 

The (instantaneous) action \eq{lingauge2} should be stationary also \wrt variations in the ratio $\la_{3}/\la_{8}$ which leaves $\la$ in \eq{lamdef} invariant. Using a lagrange multiplier $\lambda$, the extremum of the \order{g} term in \eq{lingauge2} for quark color $C=1$,
\beq\label{sint1}
S_{int}^{C=1} = \frac{g}{6}\left(\la_{3}^2+\inv{\sqrt{3}}\la_{8}^2\right)|\xv_{1}-\xv_{2}|+\lambda(\la^4-\la^4_{3}-\la^4_{8})
\eeq
for variations of $\la_{3},\la_{8}$ and $\lambda$ gives $\la_{3}^2/\la_{8}^2=\sqrt{3}$ and thus
\beq
S_{int}^{C=1} = \frac{g\la^2}{3\sqrt{3}}|\xv_{1}-\xv_{2}|
\eeq
The calculation and result is the same for $C=2$, whereas for $C=3$ the extremum of
\beq
S_{int}^{C=3} = \frac{g\la_{8}^2}{3\sqrt{3}}|\xv_{1}-\xv_{2}|+\lambda(\la^4-\la^4_{3}-\la^4_{8})
\eeq
is obtained for $\la_{3}=0,\ \la_{8}=\la$, giving $S_{int}^{C=3}=S_{int}^{C=1}\equiv S_{int}$. The fact that the stationary value of the interference term is independent of quark color is a consequence of the color singlet nature of the action \eq{lingauge2} and the color covariance of the EOM.

Having determined the parameters $\la_{a}$ and $\chat_{a}$ in the $A_{a}^0$ potential \eq{a0qcd} for each Fock state we may now proceed to impose a stationary time dependence on the bound state, as already indicated in \eq{bsqcd}. Analogously to the non-relativistic case \eq{ffcond} we have 
\beqa\label{bsamp}
i\frac{d\phi_{\alpha\beta}^{CD}(0;\xv_{1},\xv_{2})}{dt} = \rbra i\frac{d\psi_{d\beta}^{D\dag}(0,\xv_{2})}{dt} \psi_{u\alpha}^{C}(0,\xv_{1})\ket{E,0} &+& i\psi_{d\beta}^{D\dag}(0,\xv_{2})\frac{d\psi_{u\alpha}^{C}(0,\xv_{1})}{dt}\ket{E,0}\nn\\
 +\rbra\psi_{d\beta}^{D\dag}(0,\xv_{2}) \psi_{u\alpha}^{C}(0,\xv_{1})[H_{I}(0)-S_{int}]\ket{E,0} &=& E\,\phi_{\alpha\beta}^{CD}(0;\xv_{1},\xv_{2})
\eeqa
where the energy in the field contributes $-S_{int}$ as shown in \eq{phase}. The interaction hamiltonian
\beq\label{hiqcd}
H_{I}(t) = g\sum_{f=u,d}\int d^3\xv \psi_{f}^{A\dag}(t,\xv)A_{a}^0(\xv)\, T_{a}^{AB}\psi_{f}^{B}(t,\xv)
\eeq
is diagonal in color for the field \eq{a0qcd} and thus consistent with the color structure \eq{chising} of the wave function. Its matrix element in the bound state equation \eq{bsamp} with $C=D$ contributes (no sum on $C$, and neglecting terms of \order{g^2}),
\beqa\label{hicontr}
\rbra\psi_{d\beta}^{C\dag}(0,\xv_{2}) \psi_{u\alpha}^{C}(0,\xv_{1})H_{I}(0)\ket{E,0}
&=& g\sum_{a}T^{CC}_{a}\left[A^0_{a}(\xv_{1})-A^0_{a}(\xv_{2})\right]\chi(\xv_{1},\xv_{2})\nn\\
= g\sum_{a}\la^2_{a}\,|T^{CC}_{a}(\xv_{1}-\xv_{2})|\chi(\xv_{1},\xv_{2}) &=& \frac{g\la^2}{\sqrt{3}}|\xv_{1}-\xv_{2}| = 3S_{int}
\eeqa
Thus the interaction energy is independent of the quark color $C$, similarly to the instantaneous action $S_{int}$.

Using \eq{psieom} the bound state equation \eq{bsamp} for the color singlet $u\bar d$ wave function becomes
\beq\label{ffbse}
\gamma^0(-i\nv_{1}\cdot\gv +m_{u})\chi(\xv_{1},\xv_{2}) - \chi(\xv_{1},\xv_{2})\gamma^0(i\nv_{2}\cdot\gv +m_{d}) = [E-V(\xv_{1},\xv_{2})] \chi(\xv_{1},\xv_{2})
\eeq
which has the same form as \eq{ffqed} for non-relativistic QED atoms. Due to the use of the retarded vacuum \eq{retvac3} this equation may be applied also to relativistic bound states at lowest order in $\hbar$ and to \order{g} in the gauge coupling with the linear potential 
\beq\label{mespot}
V(\xv_{1},\xv_{2}) = \frac{2g\la^2}{3\sqrt{3}}|\xv_{1}-\xv_{2}|
\eeq
where $\la$ is a free parameter with dimension of mass. Separating the CM momentum $\kv$ according to
\beq\label{cmsep}
\chi(\xv_{1},\xv_{2}) = e^{i\kv\cdot(\xv_{1}+\xv_{2})/2}\,\chi_{\kv}(\xv_{1}-\xv_{2})
\eeq
the bound state equation reduces to the form \eq{bse} given in Section \ref{ffintro}.

The bound state equation \eq{ffbse} is a rather natural generalization of the Dirac equation and as such has been studied before \cite{Breit:1929zz,Geffen:1977bh,Krolikowski:1992fy,Hoyer:1985tz}. As mentioned in Section \ref{ffintro} it has several intriguing properties, in particular a correct dependence of the bound state energy $E$ on the center-of-mass momentum $\kv$, and rapid oscillations of the wave function at large distances $r$ between the quarks, where $V(r) \gg E$.

\section{Baryon bound state equation in QCD} \label{baryon}

The derivation of the baryon $uds$ equation follows the same principles as that of mesons, giving a specific three quark potential. I assume distinctly flavored quarks for simplicity.

The baryon state at $t=0$ is expressed as
\beq\label{udsket}
\ket{E,t=0} = \int\prod_{j=1}^3 d^3\yv_{j}\,\psi_{u\alpha_{1}}^{A\dag}(t=0,\yv_{1})\psi_{d\alpha_{2}}^{B\dag}(t=0,\yv_{2})\psi_{s\alpha_{3}}^{C\dag}(t=0,\yv_{3})\chi_{ABC}^{\alpha_{1}\alpha_{2}\alpha_{3}}(\yv_{1},\yv_{2},\yv_{3})\ket{0}_{R}
\eeq
where now\footnote{This definition of the retarded vacuum is different from \eq{retvac3} for mesons. This incompatibility needs to be addressed in order to treat meson-baryon interactions.}
\beq\label{retvac4}
\ket{0}_{R} = N^{-1}\prod_{\pv,\lambda,A} d_{u}^{A\dag}(\pv,\lambda)\, d_{d}^{A\dag}(\pv,\lambda)\, d_{s}^{A\dag}(\pv,\lambda) \ket{0}
\eeq
The baryon state \eq{udsket} is invariant under time independent gauge transformations $\psi^{A}(t,\xv) \to U^{AA'}(\xv)\psi^{A'}(t,\xv)$ provided the wave function is transformed as
\beq
\chi_{ABC}(\xv_{1},\xv_{2},\xv_{3}) \to U^{AA'}(\xv_{1})U^{BB'}(\xv_{2})U^{CC'}(\xv_{3})\chi_{A'B'C'}(\xv_{1},\xv_{2},\xv_{3})
\eeq
I assume that there is a gauge where the wave function has the standard color dependence
\beq\label{colbar}
\chi_{ABC}^{\alpha_{1}\alpha_{2}\alpha_{3}}(\xv_{1},\xv_{2},\xv_{3}) = \epsilon_{ABC}\chi^{\alpha_{1}\alpha_{2}\alpha_{3}}(\xv_{1},\xv_{2},\xv_{3})
\eeq
As for mesons, this allows \order{g^0} instantaneous gauge fields $A^0_{a}$ only for $a=3,8$.
The color $ABC=123$ Fock state matrix elements of the QCD equations of motion
\beq\label{bareom}
\rbra\psi_{s\alpha_{3}}^{3\dag}(t,\xv_{3}) \psi_{d\alpha_{2}}^{2\dag}(t,\xv_{2})\psi_{u\alpha_{1}}^{1\dag}(t,\xv_{1}) \Big[\partial_{\mu}F_{a}^{\mu\nu}+gf_{abc}F_{b}^{\mu\nu}A_{\mu}^c - g\sum_{f=u,d,s}\bar\psi_{f}^{A}\gamma^\nu T_{a}^{AB}\psi_{f}^{B}\Big]\ket{E,t} =0
\eeq
give for $a=3,8$ and $\nu=0$
\beq\label{bareom2}
-\nv^2 A_{a}^{0}(\xv) +gf_{abc}F_{b}^{j0}A_{c}^0 = g\sum_{j=1}^3 T_{a}^{jj}\delta^3(\xv-\xv_{j}) \hspace{1cm} (a=3,8)
\eeq
The commutator term $\propto f_{abc}$ does not contribute at \order{g} when the \order{g^0} fields appear only in the commuting elements ($b,c = 3,8$) of SU(3). The solution including homogeneous linear terms are then to \order{g}
\beqa\label{a0bar}
A^0_{3}(\xv;\{\xv_{i}\},ABC=123) &=& \la^2_{3}\,\chat_{3}\cdot\xv+\frac{g}{4\pi}\inv{2}\left(\inv{|\xv-\xv_{1}|}-\inv{|\xv-\xv_{2}|}\right)\\
A^0_{8}(\xv;\{\xv_{i}\},ABC=123) &=& \la^2_{8}\,\chat_{8}\cdot\xv+\frac{g}{4\pi}\inv{2\sqrt{3}}\left(\inv{|\xv-\xv_{1}|}+\inv{|\xv-\xv_{2}|}-2\inv{|\xv-\xv_{3}|}\right)\nn
\eeqa
This gives the instantaneous action corresponding to \eq{lingauge2},
\beq\label{lingauge3}
-\inv{4}\sum_{a}\int d^3\xv F_{\mu\nu}^{a}F^{\mu\nu}_{a} = \sum_{a=3,8}\left[\inv{2}\la_{a}^4\int d^3\xv + S_{int}^{123}+\morder{g^2}\right]
\eeq
Similarly to the meson case the \order{g^0} term proportional to the (infinite) volume of space must be universal, implying the constraint \eq{lamdef}. The \order{g} interference term for the color component $ABC=123$ is
\beqa\label{sint123}
S_{int}^{123} &=& \frac{g\la_{3}^2}{6}\,\chat_{3}\cdot(\xv_{1}-\xv_{2})+\frac{g\la_{3}^2}{6\sqrt{3}}\,\chat_{8}\cdot(\xv_{1}+\xv_{2}-2\xv_{3})\nn\\
&=&\frac{g\la_{3}^2}{6}\,|\xv_{1}-\xv_{2}|+\frac{g\la_{3}^2}{6\sqrt{3}}\,|\xv_{1}+\xv_{2}-2\xv_{3}|+\lambda(\la^4-\la^4_{3}-\la^4_{8})
\eeqa
where in the second line I used stationarity of $S_{int}^{123}$ to fix the directions of the unit vectors, $\chat_{3}\parallel \xv_{1}-\xv_{2}$ and $\chat_{8}\parallel \xv_{1}+\xv_{2}-2\xv_{3}$, and added the lagrange multiplier for the constraint \eq{lamdef}. The extremum of $S_{int}^{123}$ is obtained with
\beq
\frac{\la_{3}^2}{\la_{8}^2} = \sqrt{3}\,\frac{|\xv_{1}-\xv_{2}|}{|\xv_{1}+\xv_{2}-2\xv_{3}|}
\eeq
giving
\beq\label{sintbar}
S_{int}^{123}=\frac{g\la^2}{3\sqrt{3}}\,\sqrt{\xv_{1}^2+\xv_{2}^2+\xv_{3}^2-\xv_{1}\cdot\xv_{2}-\xv_{2}\cdot\xv_{3}-\xv_{3}\cdot\xv_{1}}
\eeq
This expression is fully symmetric under permutations of $\xv_{1}\leftrightarrow\xv_{2}\leftrightarrow\xv_{3}$, ensuring that the same result will be obtained for all color components of the wave function \eq{colbar}: $S_{int}^{123}=S_{int}^{213}= \ldots \equiv S_{int}\,$.

The stationarity condition for the baryon state \eq{udsket},
\beq
i\frac{d}{dt}\ket{E,t} = E\ket{E,t}
\eeq
imposes an \order{g} condition on the wave function $\chi_{ABC}^{\alpha_{1}\alpha_{2}\alpha_{3}}(\xv_{1},\xv_{2},\xv_{3})$ which is analogous to \eq{bsamp} for mesons. In the interaction Hamiltonian \eq{hiqcd} only the linear, \order{g^0} terms in the gauge fields \eq{a0bar} need be considered, with the parameters $\chat_{3},\ \chat_{8}$ and $\la_{3}/\la_{8}$ determined as above by the extremum of the action for each Fock state. The stationarity condition is diagonal in color and for the $ABC=123$ color component reads
\beq\label{barbse}
\sum_{j=1}^3\left[\gamma^0(-i\nv_{j}\cdot\gv +m_{j})\right]\chi(\xv_{1},\xv_{2},\xv_{3}) +g \sum_{j=1}^3 \sum_{a=3,8} T_{a}^{jj} A_{a}^0(\xv_{j})\chi = (E+S_{int})\chi
\eeq
where the interaction term on the \lhs is
\beq
\frac{g\la_{3}^2}{2}\,\chat_{3}\cdot (\xv_{1}-\xv_{2}) + \frac{g\la_{8}^2}{2\sqrt{3}}\,\chat_{8}\cdot (\xv_{1}+\xv_{2}-2\xv_{3}) = 3S_{int}(\xv_{1},\xv_{2},\xv_{3})
\eeq
The fact that $S_{int}$ given by \eq{sintbar} is a symmetric function of the quark positions $\xv_{j}$ implies that the potential is the same for all color components and thus compatible with the color structure \eq{colbar} of the wave function. The bound state equation for the $uds$ baryon wave function $\chi(\xv_{1},\xv_{2},\xv_{3})$ is then
\beq\label{barbse2}
\sum_{j=1}^3\left[\gamma^0(-i\nv_{j}\cdot\gv_{j} +m_{j})\right]\chi = (E-V)\chi
\eeq
where $\nv_{j} = \partial/\partial\xv_{j}$, the Dirac matrices $\gv_{j}$ multiply the $j$th Dirac index on $\chi^{\alpha_{1}\alpha_{2}\alpha_{3}}$, $m_{1}=m_{u}$, $m_{2}=m_{d}$, $m_{3}=m_{s}$ and the potential may be expressed as
\beq\label{barpot}
V(\xv_{1},\xv_{2},\xv_{3})=\frac{\sqrt{2}g\la^2}{3\sqrt{3}}\,\sqrt{(\xv_{1}-\xv_{2})^2+(\xv_{2}-\xv_{3})^2+(\xv_{3}-\xv_{1})^2}
\eeq
In the limit where two quarks are in the same position, \eg, $\xv_{2} = \xv_{3}$, this potential coincides with the meson potential \eq{mespot}. It is perhaps significant that a potential for more than three quarks cannot be constructed analogously, as there are then more independent quark separations than diagonal SU(3) generators.

\section{Discussion} \label{disc}

I presented an approach to gauge theory bound states that uses hamiltonian time evolution in Coulomb gauge at lowest order in $\hbar$, \ie, in the absence of loop corrections. The $\hbar\to 0$ limit selects gauge field configurations for which the action is stationary. At each instant of time, and for each Fock component of the wave function, $A^0(\xv)$ is determined by the positions of the charged constituents. The absence of loops (Born level) allows the use of retarded boundary conditions. This avoids the explicit appearance of higher Fock states arising from time ordering ($Z$-graphs), without affecting the energies of the bound states.

The field equations determine $A^0(\xv)$ only up to a homogeneous solution, which allows adding a linear term to the Coulomb potential. Such a term would not be acceptable in the form of a fixed external potential, since it would break rotational invariance and imply a field extending to spatial infinity. However, $A^0(\xv)$ is a non-propagating field which is determined by the positions of the charged constituents at each instant of time. Stationarity of the action requires a linear potential term to be oriented along the charge separation in each Fock component, ensuring rotational invariance. For neutral states the coherent sum of the contributions to $A^0(\xv\to\infty)$ from the various Fock states cancels, thus avoiding a long-range field external to the bound state. 

A linear potential implies confinement and is thus of obvious interest for QCD. The present semi-classical approach does not predict its magnitude nor why such a term should be present in QCD but not in QED. As seen from \eq{lingauge} a linear term gives a contribution to the instantaneous action which is proportional the volume of space, and thus appears as a boundary condition related to vacuum structure which is motivated by data. Its derivation from theory would require methods beyond the semi-classical approximation, such as lattice gauge calculations or the use of Dyson-Schwinger equations \cite{Roberts:2007ji,Alkofer:2009dm}.

In a purely Coulombic bound state the constituent velocity scales as $v \sim \alpha$. Hence relativistic motion $v \simeq 1$ requires a large coupling $\alpha \gsim 1$, for which the use of perturbative methods cannot be justified. A linear potential of strength $g\la^2$ dominates, in a perturbative sense, gluon exchange interactions of \order{g^2}. This enables the study of relativistic bound states even for small values of the gauge coupling $g$. The phenomenological success of the Quark Model, which uses a combination of a linear and an \order{\as} perturbative potential, indicates that the strong coupling may freeze at a sufficiently small value to justify the use of a perturbative expansion even for hadrons.     

Symmetries of the lagrangian are maintained at each order of $\hbar$ and $g$. The frame dependence of wave functions defined at equal time of the constituents is non-trivial, since boosts are dynamical operators which do not commute with the hamiltonian. In the case of a purely linear potential the energy eigenvalues $E$ of the bound state equation \eq{bse} were found \cite{Hoyer:1985tz} to have the correct dependence on the CM momentum $\kv$. The wave functions are related by a generalized Lorentz contraction \eq{sdef}, with a length scale proportional to the inverse canonical energy $1/(E-V)$ instead of $1/E$. To my knowledge this is the only case where an explicit relation between equal-time wave functions in different frames has been established, and it deserves further study. It may allow to relate spherically symmetric wave functions in the rest frame with the infinite momentum frame wave functions corresponding to quantization on the light-cone. The latter wave functions are invariant under boosts but defined \wrt a fixed direction in space \cite{Brodsky:1997de}.

The wave functions have a rapidly oscillating phase $\sim\exp(irV(r)/4)$ at distances $r$ between the constituents where the linear potential $V(r) \gg E$. Hence the relative momentum between the bound state constituents can be large, allowing support for quark distributions at low $x$ in DIS, as well as Regge type quark exchange between hadrons scattering at high energies. This is interesting also in view of the linear Regge trajectories found in the spectrum \cite{Geffen:1977bh}.

The meson and baryon solutions found here are essentially equivalent to those in \cite{Hoyer:1983xh}, which relied on some {\it ad hoc} assumptions. The present framework is based on a power expansion in $\hbar$ and $g$ and thus close to a systematic derivation of bound states from the lagrangian. This should allow to address further issues, checking the viability of the method presented here. Many open questions obviously remain. A central outstanding issue is whether spontaneous chiral symmetry breaking can be described at the semi-classical level, \eg, using a chirally non-invariant vacuum state. It is also essential to demonstrate that loop corrections can be systematically evaluated.

\vspace{1cm}
\noindent {\bf Acknowledgments}

\vspace{.2cm}

My special thanks are due to Stan Brodsky for sharing his insights into bound state structure and for continuously inspiring discussions. I have enjoyed helpful discussions also with Jeppe Andersen, Poul Damgaard, Dennis Dietrich, Stan Glazek, Matti J\"arvinen, Lorenzo Magnea, St\'ephane Peign\'e, Jian-Wei Qiu, Aleksi Vuorinen, Patta Yogendran and Feng Yuan. I am grateful for the hospitality of the CERN Theory Division, ECT* (Trento) and CP$^3$ (Odense) during the completion of this work. I have benefitted from travel support from the Magnus Ehrnrooth foundation. 

\vfill\break

\appendix

\centerline{\Large \bf Appendix}

\section{Time development using the functional integral}

In this paper I derived bound state equations using a hamiltonian method with boundary conditions specified by the retarded vacuum \eq{retvac}. The time development of fermion states may equivalently be considered in the functional integral formulation.
The boundary condition is then imposed through vacuum wave functionals $V(t_{i}),V(t_{f})$ at the initial and final times \cite{Hoyer:1982ni}, 
\beq\label{funintt}
\bra{f}{\ocal}\ket{i}= \int_{t_{i}}^{t_{f}}\dcal[\psi,\bar\psi]\,V^\dag(t_{f})\,\ocal\,V(t_{i})\exp[iS/\hbar]
\eeq
The Feynman propagator $S_{F}(x-y)$ is obtained for $\ocal=\psi(x)\bar\psi(y)$ using a wave functional $V_{F}$ with a Dirac sea of filled negative energy states \ie, the usual perturbative vacuum,
\beqa\label{feynfun}
V_{F}(t)
&=& \exp\left[-\int \dpv\inv{2E_{p}} \bar\psi(t,\pv) (\pv\cdot\gv+m) \psi(t,\pv)\right]\\
 &=&  \exp\left\{-\inv{2}\int \dpv \sum_{\lambda}\left[\psi_{+,\lambda}^\dag(t,\pv) \psi_{+,\lambda}(t,\pv) - \psi_{-,\lambda}^\dag(t,\pv) \psi_{-,\lambda}(t,\pv)\right]\right\}\nn
\eeqa
In the second expression the wave functional was diagonalized using the positive and negative projections
\beq
\psi^\alpha(t,\pv) = \sum_{\pm,\lambda}U^\alpha_{\pm,\lambda}\psi_{\pm,\lambda}(t,\pv)\ ;\hspace{1cm} U^\alpha_{+,\lambda}=\frac{u^\alpha_{\lambda}(\pv)}{\sqrt{2E_{p}}}\ ;\hspace{1cm} U^\alpha_{-,\lambda}=\frac{v^\alpha_{\lambda}(-\pv)}{\sqrt{2E_{p}}}
\eeq
where the $\pm$ refers to positive/negative energy and $\lambda=\pm\halft$ to the spin states. The minus sign in front of $\psi_{-,\lambda}^\dag(t,\pv) \psi_{-,\lambda}(t,\pv)$ in \eq{feynfun} signifies that the negative energy states are filled. This distinction between positive and negative energy field components implies that $V_{F}$ is non-local in coordinate space.

The retarded propagator $S_{R}(x-y)$ \eq{sr} is obtained using the wave functional $V_{R}$ that has unfilled negative energy states,
\beqa\label{retfun}
V_{R}(t) &=& \exp\left\{-\inv{2}\int \dpv \sum_{\lambda}\left[\psi_{+,\lambda}^\dag(t,\pv) \psi_{+,\lambda}(t,\pv) + \psi_{-,\lambda}^\dag(t,\pv) \psi_{-,\lambda}(t,\pv)\right]\right\}\nn\\
\hspace{1cm}\nn\\ 
&=& \exp\left[-\inv{2}\int d^3\xv \bar\psi(t,\xv) \gamma^0 \psi(t,\xv)\right]
\eeqa
and hence is local in $\xv$.

The functional integral method is well suited for studying the $\hbar\to 0$ limit since the $\exp[iS/\hbar]$ factor in \eq{funintt} restricts the action $S$ to a stationary value, thus fixing the gauge field at its classical value for a given charge distribution.

\end{document}